\documentclass[11pt]{article}
\usepackage{aas2pp4,graphicx}

\begin{document}

\title{Testing Modified Newtonian Dynamics with Low Surface Brightness Galaxies --- Rotation curve fits}
\author{W.J.G.~de~Blok$^1$ and S.S.
  McGaugh$^2$}
\affil{$^{1}$Kapteyn Astronomical Institute, P.O.~Box 800,
  9700 AV Groningen, The Netherlands\\ $^{2}$Department of Terrestrial
  Magnetism, Carnegie Institution of Washington, 5241 Broad Branch
  Road NW, Washington, DC 20015, USA}
\altaffiltext{1}{Present Address: Astrophysics Groups, School of Physics,
University of Melbourne, Parkville, Victoria 3052, Australia}
\altaffiltext{2}{Present Address: Physics and Astronomy, Rutgers University,
136 Frelinghuysen Road, Piscataway, NJ 08854-8019}

\begin{abstract}
We present MOND fits to 15 rotation curves of LSB galaxies.  Good fits
are readily found, although for a few galaxies minor adjustments to
the inclination are needed. Reasonable values for the stellar
mass-to-light ratios are found, as well as an approximately constant
value for the total (gas and stars) mass-to-light ratio. We show that
the LSB galaxies investigated here lie on the one, unique Tully-Fisher
relation, as predicted by MOND.  The scatter on the Tully-Fisher
relation can be completely explained by the observed scatter in the
total mass-to-light ratio. We address the question of whether MOND can
fit any arbitrary rotation curve by constructing a plausible fake
model galaxy. While MOND is unable to fit this hypothetical galaxy, a
normal dark halo fit is readily found, showing that dark matter fits
are much less selective in producing fits.  The good fits to rotation
curves of LSB galaxies support MOND, especially as these are galaxies
with large mass discrepancies deep in the MOND regime.
\end{abstract}


\section{Introduction}

The inability of the visible mass components in disk galaxies to
account for the observed rotation curves is usually interpreted as
evidence for the existence of an additional, invisible mass component.
Other theories suggest that this mass discrepancy is an indication of
a breakdown of classical Newtonian dynamics.  It is difficult to
evaluate these theories, as only a few make specific and testable
predictions.
 
One of the exceptions is the Modified Newtonian Dynamics (MOND),
advocated by Milgrom (1983, 1989) and Sanders (1990, 1996).  This
theory postulates that Newton's Law of Gravity should be modified for
very small accelerations, with the result that any need for dark
matter disappears.  Fits to rotation curves of HSB galaxies using MOND
are of equal quality as the fits made using a dark matter halo (see
Sanders 1996). MOND is however also able to satisfactorily explain
observations of the dynamics of e.g.\ dwarf galaxies and dwarf
spheroidals (see the discussion in Milgrom 1995, and also McGaugh \&
de Blok (1998b) [hereafter Paper II]).

For a complete description of MOND, its predictions, and observational
results we refer to Milgrom (1983, 1989), Sanders (1990), Begeman,
Broeils and Sanders (1991), Bekenstein \& Milgrom (1984) and Sanders
(1996).  An extensive description of MOND results in the context of
LSB galaxies is given in Paper II.

MOND assumes that the force law changes from the conventional
Newtonian form when the acceleration of a test particle is much
smaller than a limiting acceleration $a_0$, where $a_0$ is a universal
constant.  Thus, while the normal Newtonian acceleration $g_N =
GM/r^2$ which a mass $M$ exerts on a test particle at distance $r$ is
identical to the true test-particle acceleration $g$ for accelerations
$g \gg a_0$, in the MOND limit (i.e., $g \ll a_0$) the implied
Newtonian acceleration is related to the true test-particle
acceleration $g$ by $g_N = g^2/a_0$.

The acceleration $a_0$ is a fundamental parameter in the MOND theory.
From rotation curve fitting to high-quality rotation curves, Begeman
et al.\ (1991) determined a value of $1.2 \times 10^{-10}$ m s$^{-2}$
(for $H_0 = 75$ km s$^{-1}$ Mpc$^{-1}$, which we adopt throughout this
paper).

As described in Paper II, LSB galaxies provide a strong test of MOND.
Their low surface densities imply accelerations $g < a_0$, which means
that these galaxies should be almost completely in the MOND regime.
Milgrom (1983, 1989) made a number of testable predictions on the
shapes of rotation curves, and noted that low surface density galaxies
should have slowly rising rotation curves.  This expectation of MOND
is confirmed by the observed rotation curves.  In Newtonian terms this
translates in these galaxies having large mass discrepancies (McGaugh
\& de Blok 1998, hereafter Paper I).

This brings us to one of the more pronounced differences between MOND
and classical Newtonian dynamics, which is the explanation of the
Tully-Fisher (TF) relation. As is described in detail in Paper I (see
also Zwaan et al. 1995), the fact that LSB galaxies are observed to
obey the same TF relation as HSB galaxies implies a strong coupling
between the central surface brightnesses of the disks of galaxies and
their total mass-to-light ratios (which include dark matter). Assuming
standard Newtonian dynamics this implies that LSB galaxies have a
higher total mass (within the disk radius) than HSB galaxies of the
same asymptotic velocity.  It is hard to derive this result in the
standard context without a lot of fine-tuning.

MOND {\it predicts} that all galaxies should fall on one {\it
mass}-velocity relation, which takes the form $V_{\infty}^4 = MGa_0$,
where $V_{\infty}$ is the asymptotic velocity and $M$ is the total
mass of the galaxy (that is, the mass of stars and gas).  Once the
value of $a_0$ is fixed, this relation becomes absolute and can be
tested and falsified.

We use the rotation curves of 15 LSB galaxies to do a MOND
analysis. Section~2 describes the fitting procedure.
Section~3 presents the results. In Sect.~4 we discuss whether MOND can
fit any rotation curve, and we present our conclusions in Sect.~5.

\section{Fitting LSB rotation curves with MOND}

In this paper we fit the rotation curves of the sample of 15 LSB
galaxies presented in van der Hulst et al.\ (1993), de Blok, McGaugh
\& van der Hulst (1996) and de Blok \& McGaugh (1997) using the MOND
prescription.  We refer to these papers for a description of the
properties of LSB galaxies and their rotation curves.

The rotation curves were fitted following the procedure outlined in
Begeman et al.\ (1991) and Sanders (1996).  To be consistent with the
results presented in Sanders (1996) we have assumed that the stars and
the gas are both in an infinitesimally thin disk (for our purposes
this assumption has no appreciable impact on the stellar rotation
curve --- see Broeils 1992). The Newtonian rotation curves of the
visible components (stars and H{\sc i}) were determined first. The
rotation curve of the stars was computed assuming that the
mass-to-light ratio of the stars $(M/L)_{\star}$ is constant with
radius. The rotation curve of the gas was computed using the radial
H{\sc i} surface density profile, after multiplying this by 1.4 to
take the contribution of He into account.  We ignore any molecular
gas: observations suggest that LSB galaxies contain only small amounts
of molecular gas (Schombert et al.\ 1990, Knezek 1993, de Blok \& van
der Hulst 1998).  With the Newtonian accelerations known, the MOND
acceleration can be computed (see Paper II) and the observed rotation
curves fitted using a least-squares program.

The fitting procedure has three free parameters: the distance $D$ to
the galaxy; the mass-to-light ratio of the stellar disk
$(M/L)_{\star}$; and the value of the constant $a_0$. As $a_0$ is
supposed to be a universal constant we fix it at the value determined
in Begeman et al.\ (1991).  As the LSB galaxies investigated here are
sufficiently far away that the redshift is a fair indicator of the
distance, we take the distance $D$ to be fixed at the values given in
de Blok et al.\ (1996), assuming $H_0 = 75$ km s$^{-1}$
Mpc$^{-1}$. This leaves $(M/L)_{\star}$ as the {\it only} free
parameter.  For very gas rich galaxies, there are effectively no free
parameters, as the dynamics are completely dominated by the gas
component, and the contribution of the stars can be ignored for any
reasonable value of $(M/L)_{\star}$.

\subsection{Results}

The rotation curves and MOND fits are presented in
Fig.~\ref{rotcurves} and Table~1.  Of the sample presented in de Blok
\& McGaugh (1997) we have not used F571-8 and F571-V2. F571-8 is an
edge-on galaxy, and as the fits depend crucially on the photometric
profile, the simple analysis done in de Blok et al.\ (1996) does not
suffice here (due to uncertainties in the extinction; see Barnaby \&
Thronson 1994).  For F571-V2 no photometry is available.  The results
presented in Table~1 differ slightly from those presented in
Paper~II. This is because we now fit the entire shape of the rotation
curve, rather than just the asymptotic velocity.

As can be seen in Fig.~\ref{rotcurves} a good fit to the rotation
curve was readily found for the majority of the galaxies. These fits
are shown in the two left-most columns in Fig.~\ref{rotcurves}.  The
ease with which these one-parameter fits were obtained is remarkable,
especially if one keeps in mind that in LSB galaxies the dynamics are
often dominated by the gas rather than the stars. This severely limits
the possible values for $(M/L)_{\star}$ and the freedom the MOND
recipe has to produce a good fit.

For 6 galaxies a good fit did not occur trivially.  The corresponding
``bad'' fits are shown in the third column in Fig.~\ref{rotcurves}.
Were we to stop here, we would have to claim to have falsified MOND in
spite of many good fits and confirmed predictions.  It is therefore
necessary to examine the uncertainties in the data before
making such a radical step.

For each galaxy there are two parameters which might affect our
ability to obtain a fit. One is the distance to the galaxy $D$ and the
second is the inclination of the galaxy $i$ ($a_0$ is supposed to be a
universal parameter, and varying it from galaxy to galaxy would be
against the spirit of MOND).

\subsubsection{Distance Uncertainties}

We have made two-parameter fits to the 6 ``bad'' curves with
$(M/L)_{\star}$ and $D$ free. The results are shown in Table 2.
Expect for UGC 5999 we find that a two-parameter fit prefers smaller
distances which are typically $\sim 60$ per cent of the original
distance. As the galaxies are not near Virgo and have a sufficiently
large redshift that this is a fair indicator of the distance, we think
it is unlikely that these fits show a realistic picture.

A closer study of the fits shows that the distance factor is almost
entirely caused by the fit overestimating the rotation velocity in the
outer parts (cf. third column in Fig.~\ref{rotcurves}).  This is
caused by the gas-rotation curve having relatively high velocities in
the outer parts. In order to accomodate the rotation curve of the
stellar disk, the fitting procedure tries to reduce the amplitude of
the gas-rotation curve by reducing the gas mass, i.e., it brings the
galaxy in closer.

With $D$ free reasonable values of $(M/L)_{\star}$ between 0.5 and 3
are obtained.  However, one should keep in mind that the amplitude of
the gas-rotation curve in the outer parts essentially forces us to
adopt a particular value of $D$, and consequently also forces us to
adopt a value of $(M/L)_{\star}$.  Therefore we cannot use $D$ to
improve the fits.

\subsubsection{Inclination Uncertainties}

\begin{figure*}
\centering
\includegraphics[width=0.75\textwidth]{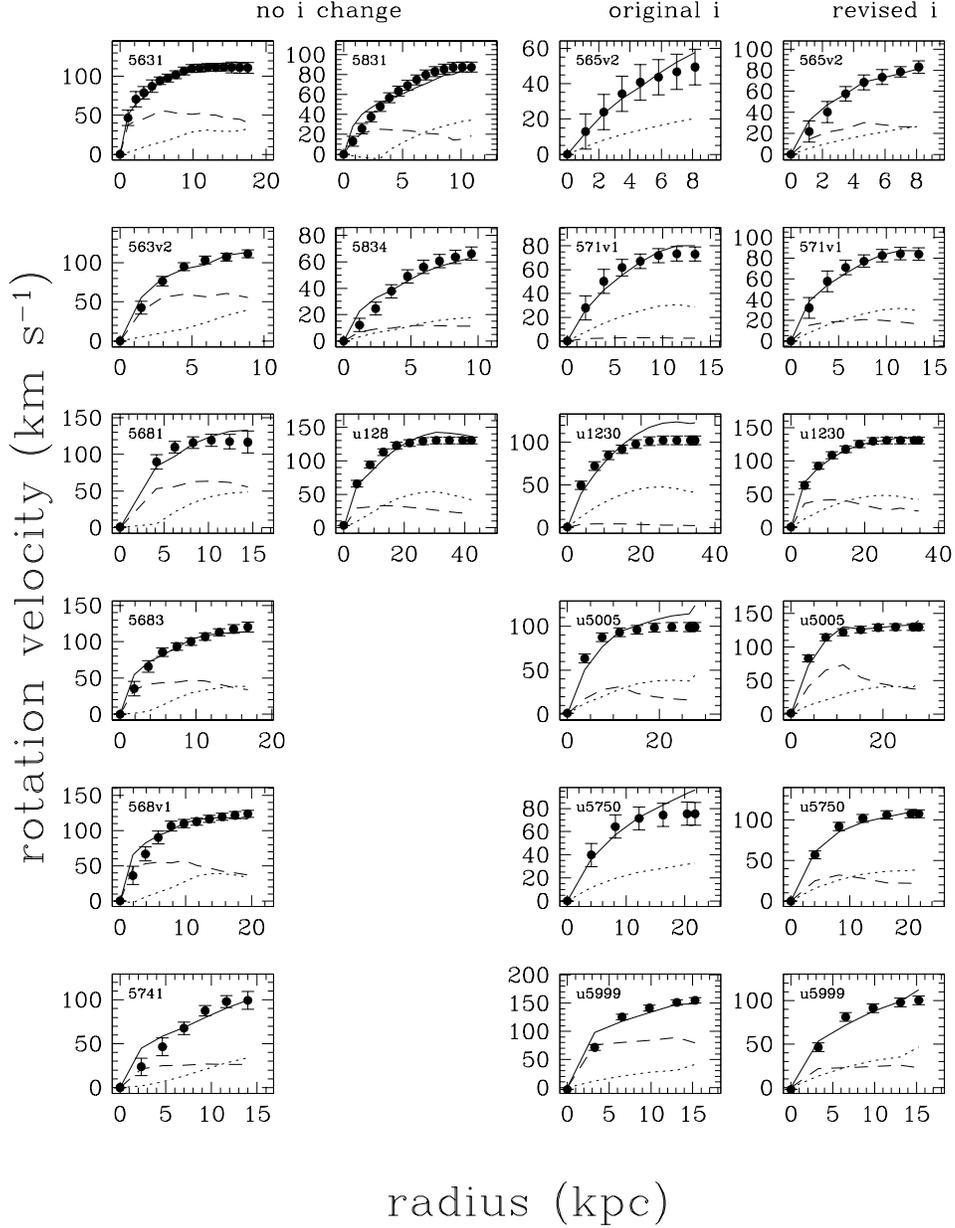}
\caption{MOND fits to the rotation curves of LSB galaxies. The two
leftmost columns show those fits where the inclination was not
adjusted. The third column shows the ``bad'' fits where the
inclination was not adjusted. The fourth, and rightmost column shows
the same 6 galaxies, but with inclinations changed to produce a good
MOND fit.  The dotted line shows the Newtonian rotation curve of the   
stellar disk, scaled with the appropriate $(M/L)_*$ value. The dotted
line
shows the Newtonian rotation curve of the gas. The solid line shows
the resulting curve derived using the MOND recipe. The solid circles
are the observed curves.}
\label{rotcurves}
\end{figure*}

The inclinations of LSB galaxies are difficult to constrain from the
shapes of their outer isophotes (McGaugh \& Bothun 1994; de Blok et
al.\ 1995, 1996).  Yet this is essentially the only information we
have which is independent of the kinematics.

The total rotation curve and the rotation curves of the stars and the
gas each depend differently on the inclination. The total rotation
velocity can be derived from the observed rotation velocity by $V_{\rm
tot} = V_{\rm obs}/\sin i$. The rotation velocity of the gas and stars
needs to be derived from the true surface density, which in turn is
determined from the observed surface density $\sigma_{\rm obs}$, so
that $V_{\rm gas,stars} \sim \sigma_{\rm obs}\cos i$.
The different dependencies on $i$ will change the ratios between the
total rotation curve and the stellar and gas rotation curves.  Changes
of only a few degrees can have quite noticeable effects for relatively
face-on galaxies.

\begin{figure*}
\centering
\includegraphics[width=0.8\textwidth]{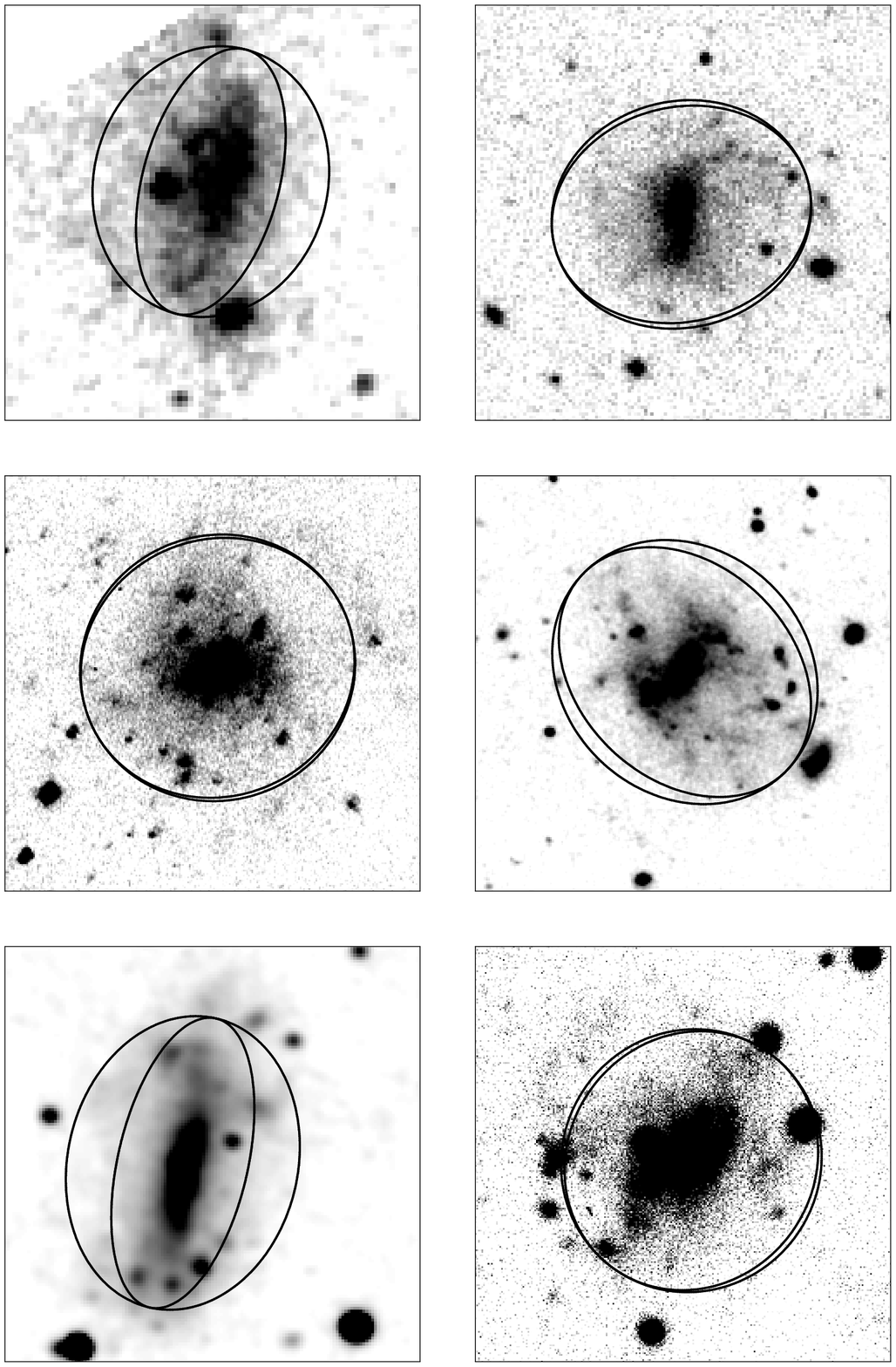}
\caption{Optical $R$-band pictures of the 6 galaxies for which the
  assumed inclination had to be changed to get a good MOND fit.  Top
  row: F565-V2 and F571-V1; Middle row: U1230 and U5005; Bottom row:
  U5750 and U5999. The superimposed ellipses show the inclinations as
  derived in de Blok et al.\ (1996) and the MOND inclinations.  In all
  cases, except for U5999, MOND prefers the lower inclination. See
  Table 1 for the exact values.}
\label{pics}
\end{figure*}

If the inclination is under-estimated, the in\-cli\-na\-tion-corrected
observed velocities are too high.  The gas contribution, which
dominates in the outer parts, has no scaling factor built in, and is
therefore unable to fit these velocities. This can be compensated for
by using a larger value of $(M/L)_{\star}$, but this will result in a
model rotation curve with the wrong shape, as it will overestimate the
velocity in the inner parts by a large factor.

The effects of over-estimating the inclination are the reverse. The
inclination-corrected observed velocities are too low, so that the
rotation curve of the gas alone suffices to fit the data in the outer
parts.  Consequently only a very modest or even zero or negative value
of $(M/L)_{\star}$ is needed. This again results in a model curve with
the wrong shape: too low velocities in the inner parts, and too high
in the outer parts.  There is only a narrow range in inclination where
both the inner and outer parts are fit equally well.

In deriving the best value for the inclination we first changed the
inclinations in a systematic way starting from the initial value from
de Blok et al.\ (1996), and rederived the mass models for each value
of the inclination. We used the $\chi^2$ value to select a range of
inclinations that yielded good fits.  This range was then narrowed by
inspecting these fits by eye, and choosing those fits that most
closely mimicked the behaviour of the high-resolution fits presented
in Sanders (1996). This means that the Newtonian rotation curve of the
stellar disk should be able to describe the innermost one-third of the
rising part of the observed rotation curve.  In practice the values
for the inclination that were determined in that way are constrained
to typically within 4 degrees.

We now discuss on a case-by-case basis the validity of changing the
inclination in these 6 galaxies.  The two right-most columns in
Fig.~\ref{rotcurves} show the original and revised fits of these two
galaxies.  Figure~\ref{pics} shows slightly smoothed optical $R$-band
pictures of the 6 galaxies with ellipses superimposed whose axis
ratios correspond to the original assumed inclination and the derived
MOND inclination.  The MOND inclination values are also given in
Table~1.

\subparagraph{F565-V2} 
At the original inclination of 60 degrees no fit with $(M/L)_{\star}>0$ could
be made. The inclination needs to be changed to 44 degrees before
positive $(M/L)_{\star}$ values are produced by the fitting procedure.  The
best fitting MOND value is 31 degrees.  This might mean that we are
missing a face-on LSB disk underlying the elongated structure we now
observe as the galaxy.  Finding such a LSB disk will be made difficult
by the presence of a bright star just outside the region shown
(cf. Fig.\ 2 in de Blok et al.\ 1996).

\subparagraph{F571-V1} The inclination had to be changed from 35 to 30
degrees, which is still fully consistent with the optical picture
(Fig.~\ref{pics}).

\subparagraph{UGC 1230} Only a very minor change was needed.
The original value derived in van der Hulst et al.\ (1993) is 22
degrees. MOND needs a value of 17 degrees in order to produce a good
fit. Note that the quality of the fit for this face-on system depends
critically on the inclination.  Using MOND to fit the rotation curve
of a face-on galaxy could actually tie down its inclination to within
a few degrees. This is impossible with a classical dark halo fit.

\subparagraph{UGC 5005}
The original value for the inclination of this galaxy is 41
degrees. MOND demands an inclination in the range 26 to 30 degrees for
the best fits. Here we adopt a value of 30 degrees. Fig.~\ref{pics}
shows that this value is still consistent with the observed
axis ratio.

\subparagraph{UGC 5750} Good solutions are only found for inclinations
between of 38 and 42 degrees. The value of 64 degrees from van der
Hulst et al.\ (1993) is ruled out. The new low value can only be
consistent with the optical axis ratio if there is a LSB face-on disk
underlying the observed bar-like structure.  This might be consistent
with the H{\sc i} column density map in van der Hulst et al. (1993)
where one can derive an inclination of $\sim 35$ degrees for the
outermost column density contours. Deeper observations of this galaxy
should be able to test this.

\subparagraph{UGC 5999} MOND demands a value for the inclination of 22 degrees instead of 14
degrees.  As shown in Fig.~\ref{pics} this results in only a very
minor change in the axis ratios.

It is clear that in most cases the changes needed in the inclination
to produce good fits are only minor.  The only two challenging objects
are F565-V2 and UGC 5750.  To justify their inclinations one needs in
both cases a more face-on disk underlying the observed part of the
galaxy.  In both cases, deeper observations should provide more
definite determinations of the inclinations. If these new values are
still inconsistent with the MOND values, then this may be a
problem for the theory.

It is striking that the fraction of barred galaxies in Figure~2 is
much higher ($\sim 50$~per cent) than in the entire LSB sample ($<
10$~per cent).  This may play some role in the inclination
uncertainties.  Also, note that the general preference is for slightly
more face-on disks.  This is expected for systematic errors introduced
by finite thickness disks or deviations from purely circular shapes.
In the rest of this paper we will adopt the new MOND inclinations and
the corresponding results, but we will distinguish them from the
``good'' fits by using different symbols.  In practice the conclusions
will not be affected.

\begin{figure}[t]
\centering
\includegraphics[width=0.9\linewidth]{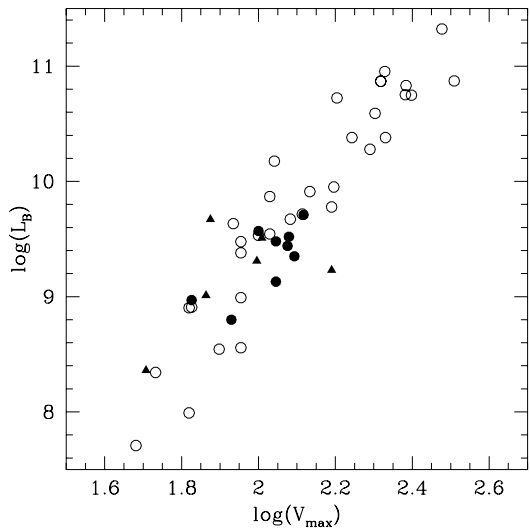}
\caption{The $B$-band Tully-Fisher relation for HSB and LSB
  galaxies. The open symbols show the data from Sanders (1996). The    
  filled circles LSB galaxies where the inclination was not
  changed. The filled triangles represent the LSB galaxies where $i$  
  had to be changed in order to get a good fit.}
\label{TF}
\end{figure}

\begin{table*}
\begin{minipage}{144mm}
\caption[]{Results of MOND fits\label{tabel1}}
\begin{tabular}{lrrrrrcccl}
\noalign{\vskip 2pt}
\tableline
 Name & $V_{\rm max}$ & $L_B$ &$\mu_B$ &$D$ &$M_{\rm gas}$ &
 $\left(\frac{M}{L_B}\right)_{\star}$ & $\left(\frac{M}{L_R}\right)_{\star}$
 &$\left(\frac{M_{\rm tot}}{L_B}\right)$ &$i\ [i_{\rm MOND}]$\\
\tableline
 F563-1  & 111  &0.135  &23.6  &45  &0.385 & 2.97 & 3.99  &5.82  &25\\
 F563-V2 & 111  &0.302  &22.1  &61  &0.321 & 1.81 &(2.38) &2.87  &29\\
 F565-V2 & 51   &0.023  &24.7  &48  &0.084 & 2.18 & 2.70  &5.85  &60 [31]\\
 F568-1  & 119  &0.275  &23.8  &85  &0.557 & 3.01 & 3.96  &5.03  &26\\
 F568-3  & 120  &0.331  &23.1  &77  &0.394 & 1.33 & 1.62  &2.52  &40\\
 F568-V1 & 124  &0.224  &23.3  &80  &0.343 & 2.96 & 3.80  &4.49  &40\\
 F571-V1 & 73   &0.102  &24.0  &79  &0.164 & 0.66 & 0.84  &2.26  &35 [30]\\
 F574-1  & 100  &0.372  &23.3  &96  &0.485 &(0.54)& 0.71  &1.85  &65\\
 F583-1  & 85   &0.063  &24.1  &32  &0.243 & 1.68 & 1.96  &5.53  &63\\
 F583-4  & 67   &0.093  &23.8  &49  &0.077 &(0.24)& 0.31  &1.07  &55\\
 U128    & 131  &0.513  &23.2  &60  &0.882 & 0.84 & 1.06  &2.56  &55\\
 U1230   & 102  &0.324  &23.3  &51  &0.812 & 1.16 & 1.71  &3.67  &22 [17]\\
 U5005   & 99   &0.204  &23.8  &52  &0.406 &(3.65)& 4.80  &5.64  &41 [30]\\
 U5750   & 75   &0.468  &23.5  &56  &0.140 &(0.68)& 0.89  &0.98  &64 [39]\\
 U5999   & 155  &0.170  &23.5  &45  &0.252 &(0.53)& 0.70  &2.01  &14 [22]\\
\tableline
\end{tabular}
\tablecomments{
Units: $V_{\rm max}$ is in km s$^{-1}$, $L_B$ in $10^{10} L_{\odot}$,
$M_{gas}$ in $10^{10} M_{\odot}$
$\mu_B$ in mag arcsec$^{-2}$, $D$ in Mpc, and $(M/L)$ in
$M_{\odot}/L_{\odot}$ in the respective passbands; $i$ is in degrees. 
Values between brackets are $B$-band values that were converted from
$R$-band observations or vice versa. Inclination values between square
brackets are values demanded by MOND to produce good fits.}
\end{minipage}
\end{table*}

\section{Results}

The main result is that MOND is able to explain the slowly-rising,
non-flat rotation curves of galaxies with low surface brightnesses,
where a large mass discrepancy is inferred in the classical, Newtonian
case.  The resulting stellar and total mass-to-light ratios enable us
to derive some additional results which we will discuss below.  In
this discussion, we will also make use of the results of the MOND fits
to the rotation curves of a large sample of HSB galaxies, presented in
Sanders (1996).

\subsection{The Tully-Fisher relation and mass-to-light ratios}

MOND predicts a unique and exact {\it mass}-velocity relation $M_{\rm
tot} \propto V^4$, where $M_{\rm tot}$ denotes the total mass, that
is, $M_{\rm tot} = M_{\star} + M_{\rm gas}$.  Converting this to more
directly observable quantities yields
\begin{equation}
V^4 = Ga_0\left(\frac{M_{\rm tot}}{L}\right)L.
\end{equation}
The exact $M-V$ relation thus translates into an identical $L-V$
relation {\it if} $M_{\rm tot}/L$ is constant. Any scatter in $M_{\rm
tot}/L$ will show up as scatter in the $L-V$ relation (that is, as
scatter in the TF relation).

MOND thus implies that there is a unique TF relation, where any
scatter must is caused by scatter in $M_{\rm tot}/L$. The small
observed scatter in the TF relation thus implies there is only a small
scatter in $M_{\rm tot}/L$. 

As discussed extensively in Papers I and II, in the classical
dark-halo TF relation the only way to keep galaxies on the TF relation
is by having a fine-tuned relation between the optical surface
brightness of a galaxy and its total mass-to-light ratio (i.e.,
including the dark matter --- Zwaan et al.\ 1995).  In MOND all that
is needed is that the scatter in $M_{\rm tot}/L$ is small enough to be
consistent with the observed scatter in the TF relation.

In Fig.~\ref{TF} we show the $B$-band MOND TF relation for HSB
galaxies as derived by Sanders (1996), with the LSB data superimposed.
It is clear that there is no systematic offset between the HSB and the
LSB galaxies. We now study the implied small scatter in $M/L$ a bit more
closely.

\begin{table}[t]
\begin{minipage}{70mm}
\caption[]{Results of MOND fits with distance free\label{tabel2}}
\begin{tabular}{lrrr}
\noalign{\vskip 2pt} 
\tableline
Galaxy  & $D$ & $D_{MOND}/D$ & $M/L_{*,df}$ \\
\tableline
F565-V2 & 48 & $0.60 \pm 0.07$ & $0.33 \pm 0.15$\\
F571-V1 & 79 & $0.65 \pm 0.11$ & $0.79 \pm 0.25$\\
U1230   & 51 & $0.63 \pm 0.03$ & $0.58 \pm 0.07$\\
U5005   & 52 & $0.53 \pm 0.03$ & $2.04 \pm 0.09$\\
U5750   & 56 & $0.45 \pm 0.55$ & $0.43 \pm 0.13$\\
U5999   & 45 & $2.82 \pm 0.39$ & $3.61 \pm 1.65$\\
\tableline
\end{tabular}
\end{minipage}
\end{table}

\begin{figure*}
\centering
\includegraphics[width=0.3\textwidth]{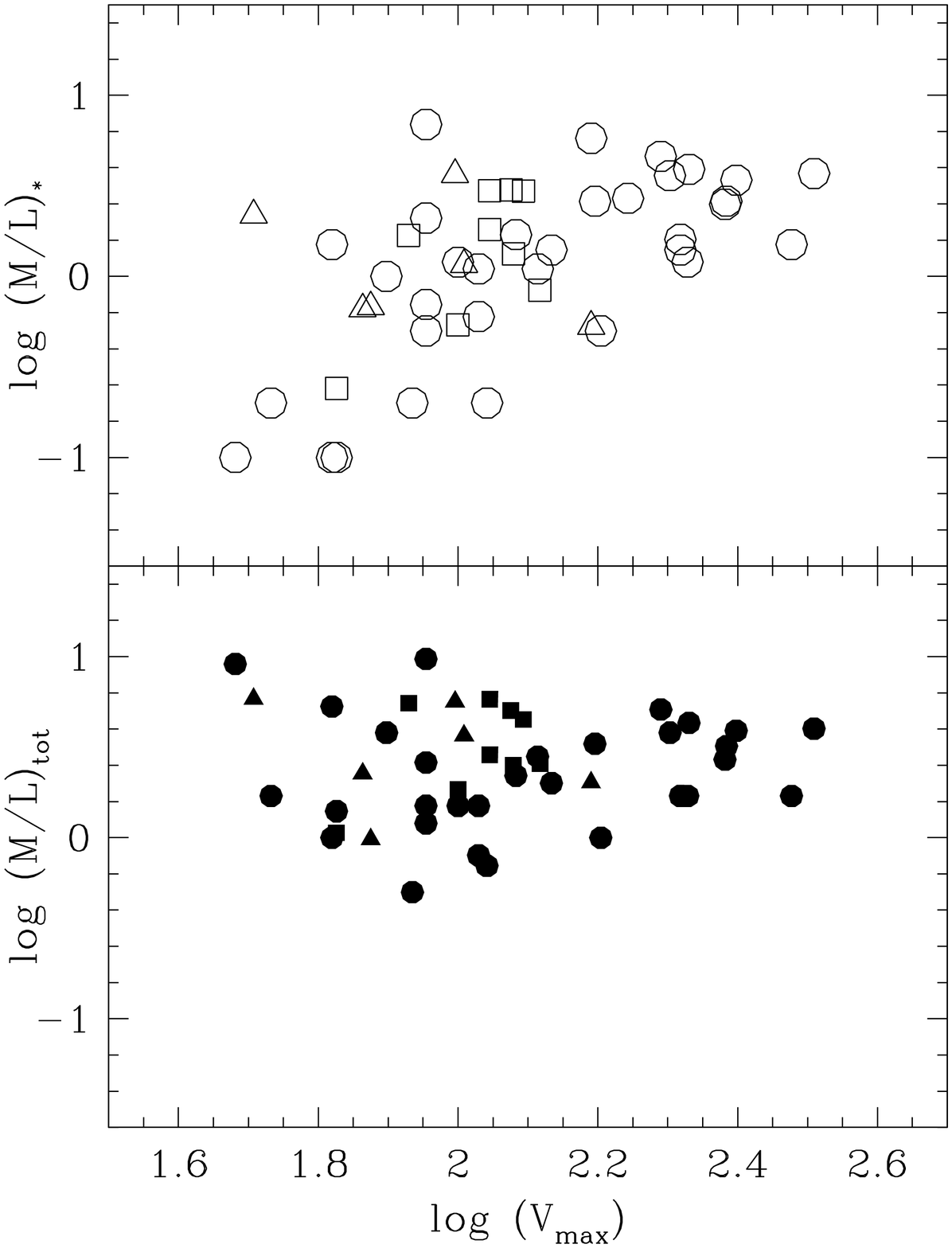}
\includegraphics[width=0.3\textwidth]{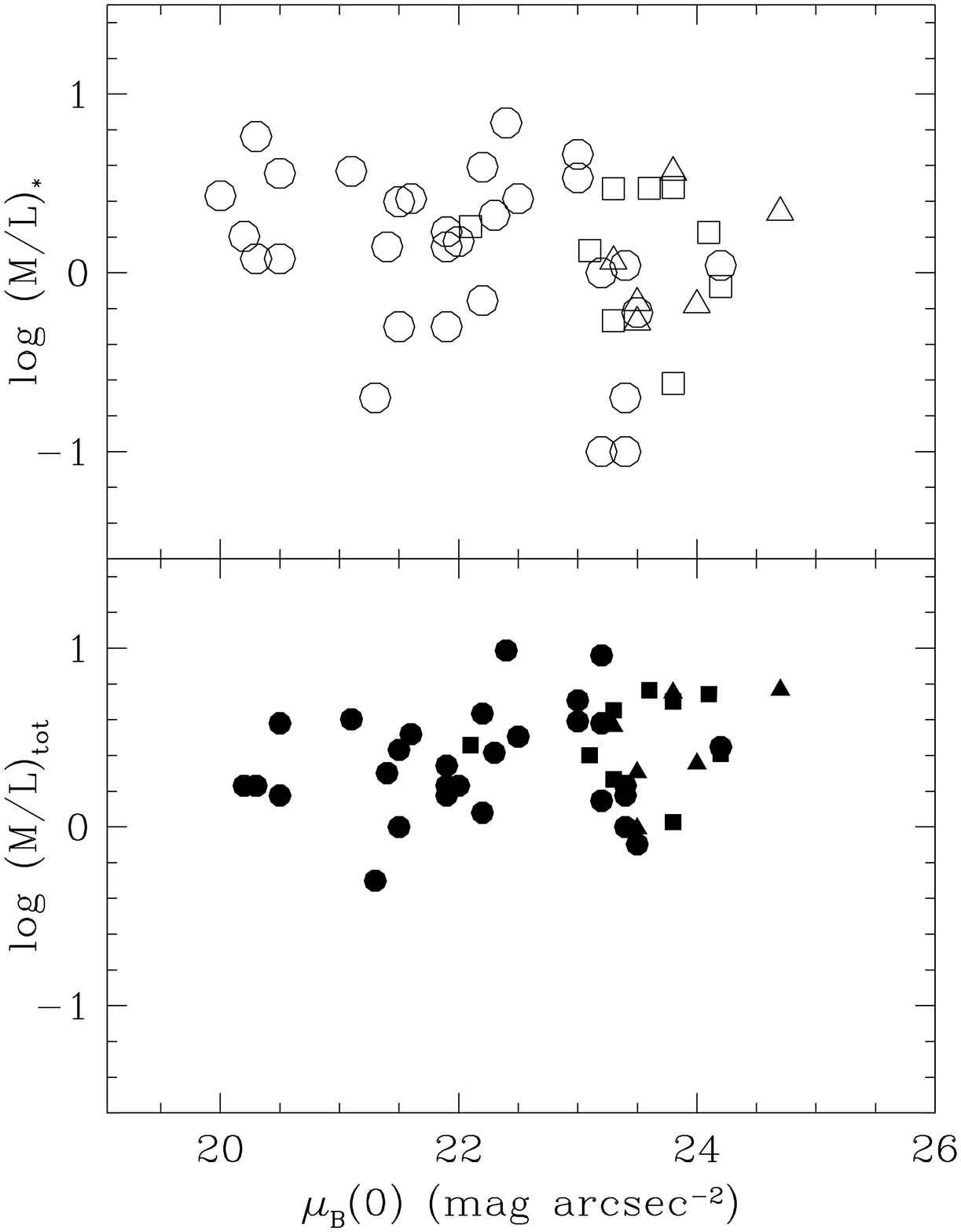}  
\includegraphics[width=0.3\textwidth]{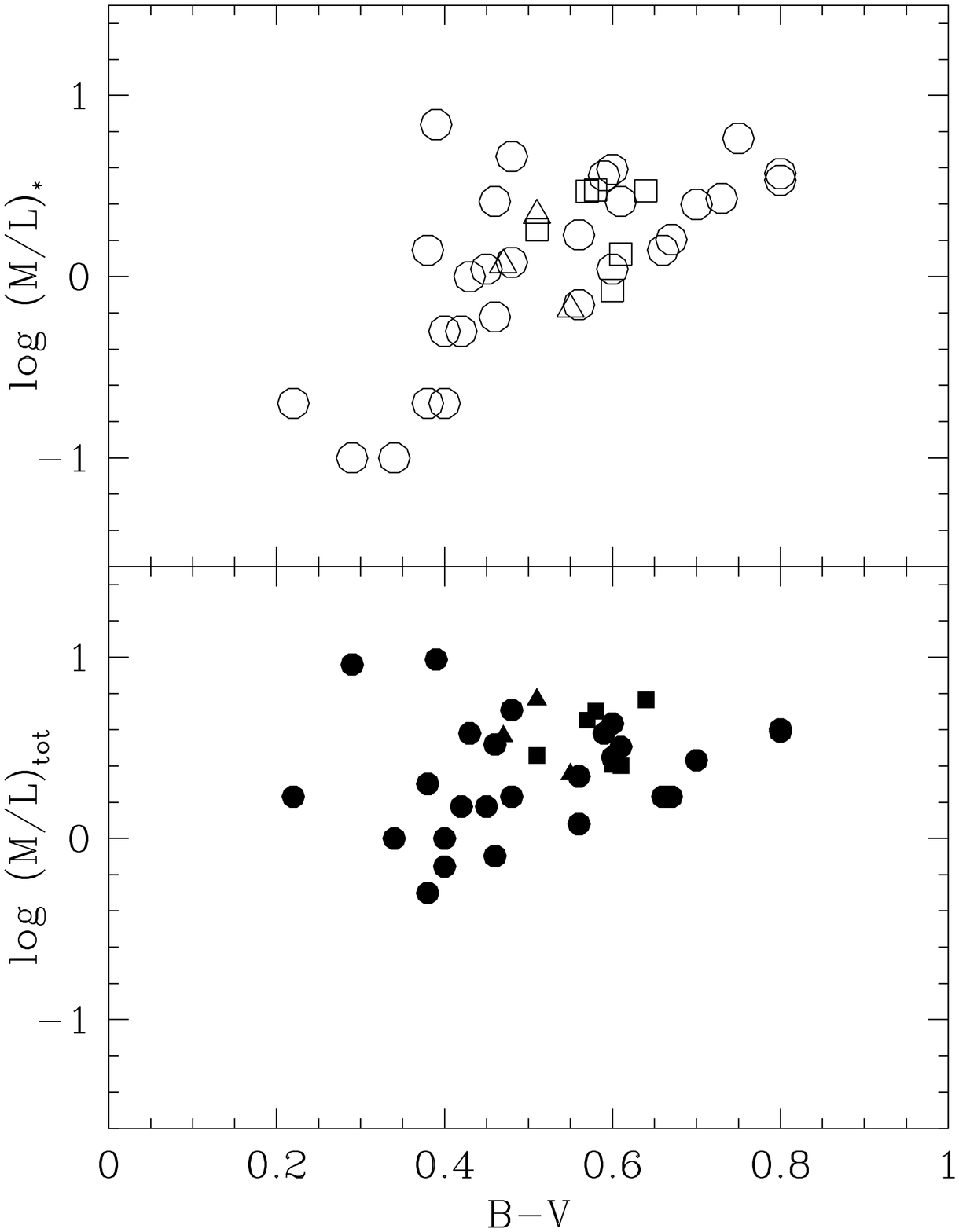}  
\caption{Mass-to-light ratios derived from the MOND fits.  Open
  symbols show the stellar mass-to-light ratio $M/L_*$; filled symbols
  show the total mass-to-light ratio $M_{\rm tot}/L$.  The circles
  represent data from Sanders (1996). The squares represent LSB   
  galaxies for which inclinations did not have to be changed. The 
  triangles represent the 6 LSB galaxies where the inclination was
  changed.  Left panel: plotted against maximum rotation velocity 
  derived from the rotation curves.  Center panel: plotted against the
  central $B$-band surface brightness. Right panel: plotted against   
  $B-V$ colour.}
\label{ML}
\end{figure*}

Figure~\ref{ML} shows the stellar mass-to-light ratio $(M/L)_{\star}$ and the
total mass-to-light ratio $M_{\rm tot}/L$ as a function of maximum
rotation velocity $V_{\rm max}$, central surface brightness $\mu_0$
and color $B-V$.

\begin{figure}[t]
\centering
\includegraphics[width=0.45\textwidth]{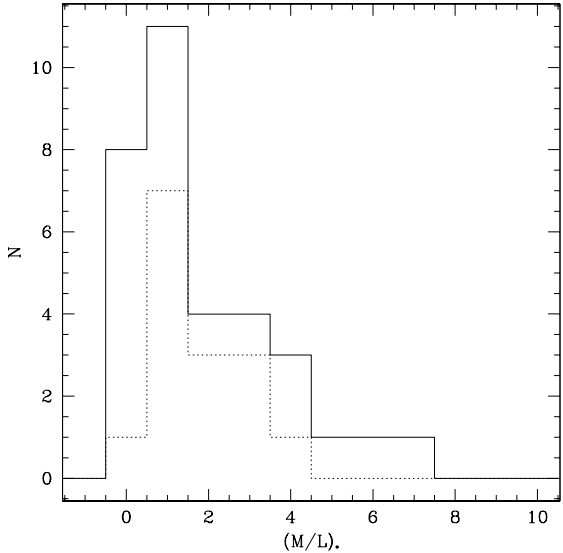}
\caption{Histogram of the values of $M/L_*$ as found from fitting of
rotation curves. The dotted histogram indicated the LSB data
superimposed on the HSB data (full line).The range and shape of both
distributions is similar. Both peak between $M/L_*$ value of 0 and 1.
The median of the HSB distribution is 1.4, that of the LSB
distribution is 1.3. Both distributions are consistent with values
normally derived from population synthesis models.}
\label{MLhist}
\end{figure}

$(M/L)_{\star}$ overall is clearly increasing with $V_{\rm max}$, consistent
with the color-magnitude relation.  The LSB galaxies are
indistinguishable from the other galaxies; a situation which is in
sharp contrast with the maximum disk results (de Blok \& McGaugh
1997).  There is no clear trend with central surface brightness,
although the scatter increases towards lower surface brightnesses,
which probably just reflects the large importance of the gas in these
galaxies and the large effects which small amounts of star formation
can have on the blue luminosity of LSB galaxies. This increase in
scatter in $(M/L)_{\star}$ is also apparent at the bluest $B-V$ colors.
The values of $(M/L)_{\star}$ are reasonable and vary between $\sim 0$
and $\sim 5$ (see Fig.~\ref{MLhist}). The distribution peaks at
$(M/L)_{\star}\sim 1$. The median of the HSB data equals 1.4, while that
of the LSB data equals 1.3.  Most fall within the range of
$(M/L)_{\star}$ predicted by population synthesis models (see Sanders
1996).  No negative values of $(M/L)_{\star}$ were found. MOND
therefore does not over-correct the mass-discrepancy. Measurements in
e.g.\ the near-infrared should yield in a much smaller range in
$(M/L)_{\star}$ values, as tentatively indicated by Sanders (1996) and
Sanders \& Verheijen (1998).

The relevant parameter for the Tully-Fisher relation is however the
{\it total} mass-to-light ratio, defined as $(M_{\rm gas}+M_{\star})/L$. This
quantity is also shown in Fig.~\ref{ML}.  A comparison with the
$(M/L)_{\star}$ data points shows that inclusion of the gas mass has made the
increasing trend with $V_{\rm max}$ disappear.  $M_{\rm tot}/L$ does not
show any systematic trend with $V_{\rm max}$, $B-V$ and $\mu_0$. The
gas mass thus makes a large contribution for low luminosity and low
surface brightness galaxies.
The average value of $M_{\rm tot}/L$ is 2.5.  The dispersion in
$\log(M_{\rm tot}/L)$ is 0.30 dex, which compares to a dispersion in
$\log L_B$ in the blue TF relation of 0.33 dex.  The scatter in the TF
relation is thus entirely consistent with, and can be explained by,
the scatter in the values of $M_{\rm tot}/L$.

The LSB galaxies in all respects confirm the predictions made by
Milgrom (1983).  These galaxies are deep in the MOND regime, with the
lowest accelerations currently known.  They provide the clearest and
strongest test of MOND made to date.  

\section{Does MOND fit any rotation curve?}

One of the myths surrounding MOND is that it is specifically designed
to fit rotation curves and is therefore guaranteed to do so.  Reality
is a bit more complicated than this.  MOND {\it predicts} what the
rotation curve will be on the basis of solely the {\it observed}
matter distribution. This is a fundamental difference from the dark
matter theory where a dark halo is fit to whatever is left after the
subtraction of the luminous component.  The dark matter theory does
not predict rotation curves, whereas in MOND there is a unique and
unambiguous correspondence between the observed matter distribution
and the rotation curve.

\begin{figure*}[t]
\centering
\includegraphics[width=0.45\textwidth]{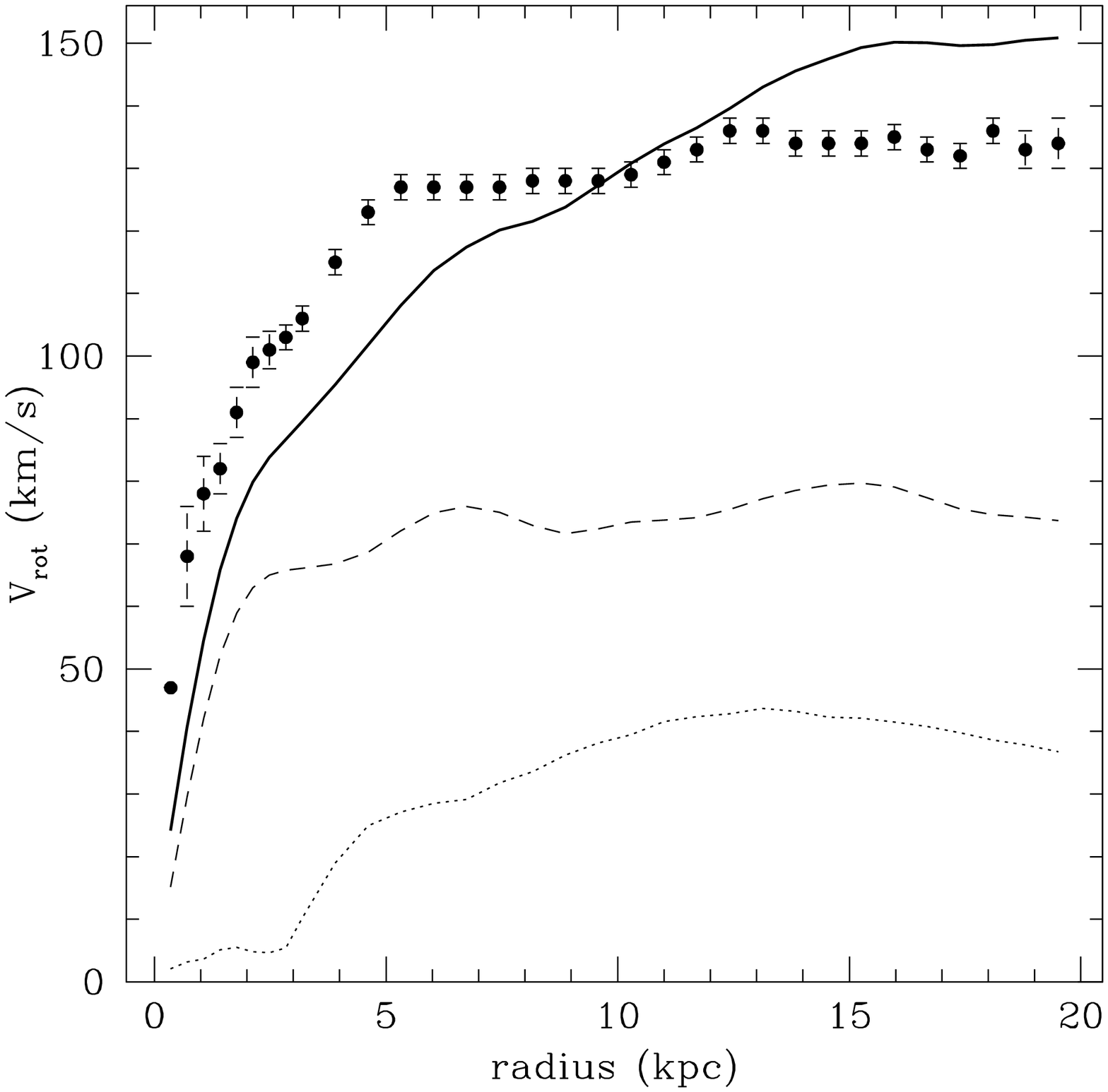}
\includegraphics[width=0.45\textwidth]{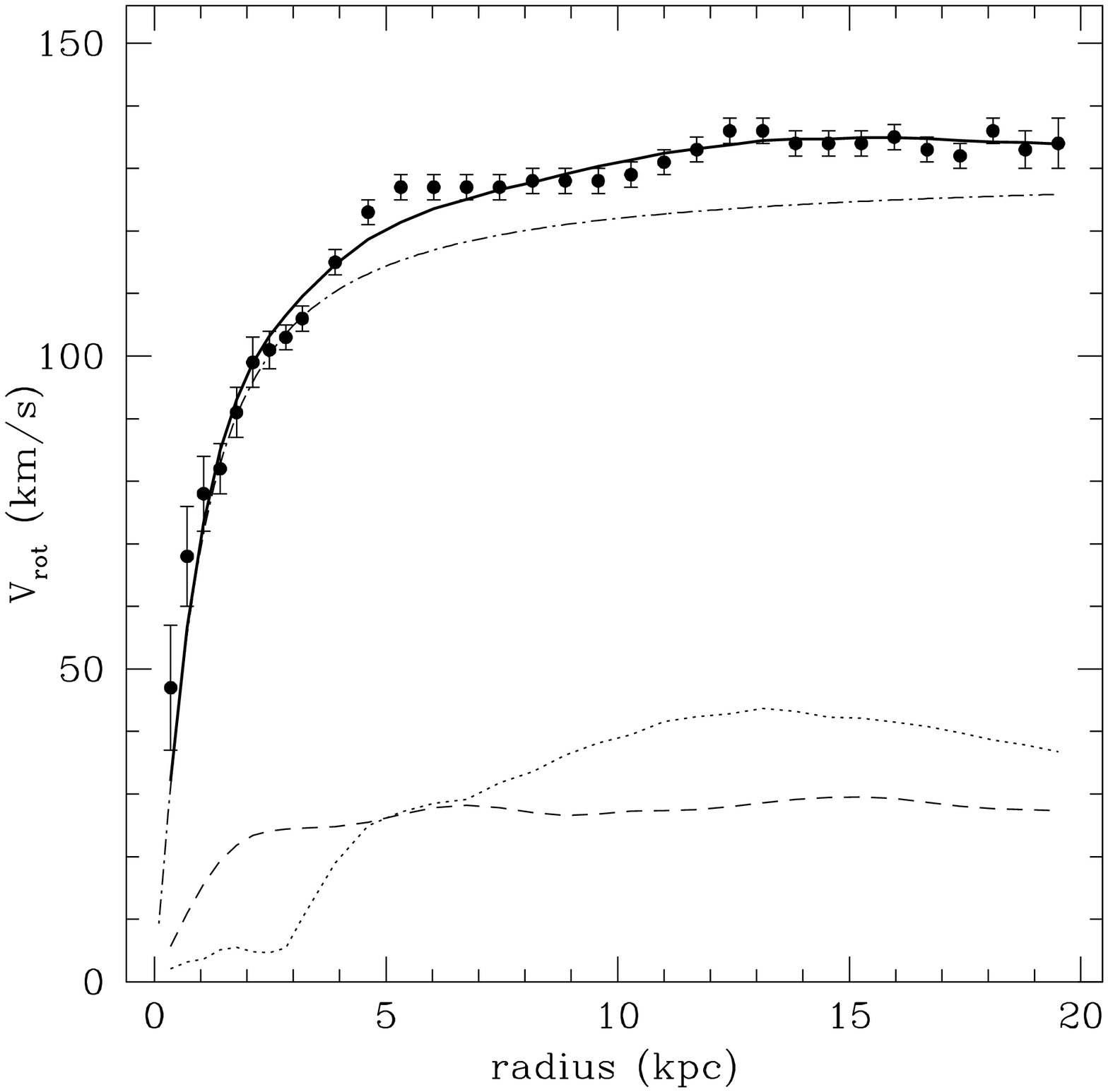}  
\caption{Fits of a hybrid NGC2403/U128 galaxy. This galaxy was
constructed by using the NGC 2403 data, but replacing its photometry
by that of UGC 128. The left panel shows the MOND result. MOND is   
unable to derive a good fit. The fit shown here has $a_0$ fixed and 
distance fixed at 3.3 Mpc. The value of $(M/L)_* = 6.0 \pm 0.5$.  The
right panel shows the classical fit with an isothermal dark matter   
halo. A good fit can be found with the following parameters: $(M/L)_*
= 0.8 \pm 0.2$, and an isothermal halo with $R_C = 0.80 \pm 0.04$ kpc
and $V_{\infty} = 130$ km s$^{-1}$ (fixed during fitting). The
corresponding core density is $\rho_0 = 487 \cdot 10^{-3}$
$M_{\odot}/$pc$^{-3}$. The dotted line shows the gas rotation curve,
the dashed line the disk rotation curve. The long dashed line in the
right panel shows the rotation curve. The heavy solid line in both  
panels is the final model rotation curve.}
\label{hybrid}
\end{figure*}

This results in MOND having much less freedom in getting the ``right''
answer than the dark matter rotation curve fits.  This is already
clear from fitting rotation curves of HSB galaxies. These show that
the inner parts of HSB galaxies are still in the Newtonian regime, and
that therefore a Newtonian (i.e., ``classical'') fit in these parts
fixes $(M/L)_{\star}$. Once $(M/L)_{\star}$ is fixed, there is no more
freedom in the fit. If the rotation curve in the outer parts does not
exactly behave in the way MOND wants it to, then no good fits can be
made.

We illustrate all of this by the following example. The HSB/LSB pair
NGC2403/UGC 128 are two galaxies that occupy identical positions on
the Tully Fisher relation, with indistinguishable luminosities and
flat rotation amplitudes, but with very different central surface
brightnesses. This is extensively described in de Blok \& McGaugh
(1996).  In the classical, dark matter picture, the fact that these
galaxies are on the same position on the TF relation means their mass
distributions must be related. Whether this means that their halos are
identical or not is unclear.

We will now construct a hybrid model galaxy, by taking the rotation
curve of NGC 2403, but where we will replace the optical photometry of
NGC 2403 by that of UGC 128. This results essentially in a low surface
brightness version of NGC 2403.  The similarity of the two galaxies
maximizes the opportunity for both MOND and dark matter theories to
achieve a fit to the hypothetical hybrid.

We have fitted this hypothetical galaxy with both the MOND and dark
matter fitting procedures. We present the results in
Fig.~\ref{hybrid}.  It is clear that in the MOND case no good fit can
be made. The observed mass distribution is not related to the rotation
curve, hence the bad fit.  The implied stellar mass-to-light ratio is
not reasonable, and both galaxies have rather high, well determined
inclinations so that adjusting the inclination within the errors will
not have an impact on the procedure.  In the dark matter case one {\it
can} reach a very tolerable fit, which is comparable in quality with
the fit of the real NGC 2403 (see Begeman, Broeils and Sanders 1991,
their Fig.~1).

The conclusion we can reach from this is that MOND can {\it not\/} fit
any arbitrary rotation curve, dispelling the myth which asserts it was
designed to do just this.  MOND was designed to produce an
asymptotically flat rotation curve, but this is no guarantee that the
application of the MOND force law to the observed luminous mass
distribution will obtain the correct shape for just any rotation
curve.  There must be a very close relation between the observed mass
distribution and the rotation curve (both of which can be determined
independently), and this relation must be the MOND recipe if one wants
to derive the latter solely from the former.

\section{Summary}

The good fits to LSB galaxies rotation curves support MOND, especially
so as these are galaxies with large mass discrepancies in their
observed disks, where the MOND effects are strongest.

The MOND fits furthermore {\it test} the MOND theory.  This is not the
case with dark matter fits, where the properties of the dark
matter discrepancy are {\it derived} from the fits. These fits {\it
define} the properties of the mass discrepancy. It is therefore not
possible to falsify nor confirm the dark matter hypothesis from
rotation curves alone, as can be done with the MOND hypothesis. We
show this by fitting a hypothetical non-physical galaxy: the MOND
theory is unable to produce a good fit (which in this case is a good
thing), whereas we can get a good fit with the dark matter theory
(showing it is more flexible in fitting non-physical models).

In principle, there are an infinite number of things the rotation
curves of galaxies could do given the presence of an invisible mass
component.  There are many plausible predictions for what they are
expected to do given various ideas about dark matter.  In the case of
MOND, there is one and only one thing rotation curves can do. This is
precisely what they do.

One last remarkable result is the great efficiency with which MOND can
describe the rotation curves of galaxies of very different types. Even
if MOND is ultimately falsified, it would still be worth knowing why
it works so well.  Until a MOND cosmology can be derived (see Sanders
1998) it remains a recipe for describing rotation curves, but as a
recipe it is far superior to the dark halo recipe, and this raises the
question of {\it why} MOND works so well.  MOND is after all only a
simple analytic formula which gets $V(R)$ correct based solely on the
luminous mass.  If the MOND phenomenology arises as the result of dark
matter, then this would imply that the dark matter must ``know'' about
both the distribution of light {\it and} the MOND formula, and arrange
itself appropriately, right down to amplifying the bumps and wiggles
in the stellar and gas rotation curves by the appropriate amount.

\end{document}